\documentclass[aps,pra,twocolumn,groupedaddress,showpacs]{revtex4}
\usepackage{graphicx}
\usepackage{amssymb}

\begin{document}

\title{Casimir effect between dissimilar materials: a test for the proximity theorem}

\author{Cecilia Noguez}
\email[Corresponding author. Email:]{cecilia@fisica.unam.mx}
\affiliation{Instituto de F\'{\i}sica, Universidad Nacional Aut\'onoma
 de M\'exico, Apartado Postal 20-364, D.F. 01000,  M\'exico}

\author{C. E. Rom\'an-Vel\'azquez}
\affiliation{Instituto de F\'{\i}sica, Universidad Nacional Aut\'onoma
 de M\'exico, Apartado Postal 20-364, D.F. 01000,  M\'exico}

\date{\today}
\begin{abstract}
We calculate exactly the Casimir force between a spherical particle and a plane, both with arbitrary dielectric properties, in the non-retarded limit. Using a Spectral Representation formalism, we show that the Casimir force of a sphere made of a material A and a plane made of a material B, differ from the case when the sphere is made of B, and the plane is made of A.  The differences in energy and force show the importance of the geometry, and make evident the necessity of realistic descriptions of the sphere-plane system beyond the Proximity Theorem approximation. 
\end{abstract}


\maketitle
\section{Introduction}
In recent years, accurate experiments~\cite{bressi,lamoraux,mohideen,chan,decca} have been performed to measure the Casimir force which is one of the macroscopic manifestations of the fluctuations of the quantum vacuum~\cite{casimir}. Most of these experiments~\cite{lamoraux,mohideen,chan,decca} have been done using a spherical surface and a plane, instead of two parallel planes, as originally proposed Casimir~\cite{casimir}. The interpretation of the Casimir force in these experiments is based on the Proximity Theorem which was developed by Derjaguin and collaborators~\cite{proximidad} to estimate the Casimir force between two curved surfaces of radii $R_1$ and $R_2$. The Proximity Theorem assumes that the force on a small area of one curved surface is due to locally ``flat'' portions on the other curved surface, such that, the force per unit area is
\[ 
\mathcal{F}(z) = 2 \pi \left(\frac{R_1R_2}{R_1 + R_2}\right) {\mathcal V}(z), \label{ptf}
\]
where $\mathcal{V}(z)$ is the Casimir energy per unit area between parallel planes separated by a distance $z$. In the limit, when $R_1 =R$, and $R_2 \to \infty$, the problem reduces to the case of a sphere of radius $R$ and a flat plane, that yields to $\mathcal{F}(z) = 2 \pi R \mathcal{V}(z).$ The force obtained using the Proximity Theorem is a monotonic function of $z$, where at ``large'' distances $\mathcal{F}(z)\propto z^{-3}$, while at short distances $ \mathcal{F}(z) \propto z^{-2}$. This theorem is supposed to hold when $z \ll R_1, R_2$, however, the validity of the Proximity Theorem approach has not been proved yet. 

Casimir showed that the energy $\mathcal{U}(z)$, between two parallel perfect conductor planes can be found through the change of the zero-point energy of the classical electromagnetic field~\cite{casimir}, like
\begin{equation}
\mathcal{U}(z) = \frac{\hbar}{2} \sum_i [\omega_i(z) - \omega_i(z\to \infty)], \label{uint}
\end{equation}
where $\omega_i(z)$ are the proper modes that satisfy the boundary conditions of the electromagnetic field at the planes which are separated a distance $z$. For real materials, Lifshitz obtained a formula to calculate the Casimir force between parallel dielectric planes~\cite{lifshitz}. The force per unit area, according to the Lifshitz formula, is 
\[ 
f(z) = \frac{\hbar c}{2\pi^2} \int_0^\infty dQQ \int_{q\ge 0} dk \frac{k^3g}{q} {\rm Re} \frac{1}{\tilde k} \left[ \frac{1}{\xi^s -1} +  \frac{1}{\xi^p -1} \right],
\] 
with $\xi^\alpha = [r^\alpha_1 r^\alpha_2\exp(2i\tilde k z)]^{-1}$, where $r_j^\alpha$ is the reflection amplitude coefficient of the plane $j$ for the electromagnetic field with  $\alpha$-polarization. Here, $j=1$, or $2$, $\alpha = s$ or $p$, $q=\omega/c$, $\omega$ is the frequency and $c$ is the speed of light. Here, $\vec{q} =(\vec{Q},k)$ is the vacuum wavevector with projections parallel to the surface $\vec{Q}$, and normal to the surface $k$, $\tilde{k} = k + i0^+$, and $g$ is the photon occupation number of the state $k$. 

The Lifshitz formula depends only on the reflection coefficients of the planes and the separation between them. This reflection coefficients can be calculated for arbitrary material using the surface impedance formalism for semi-infinite slabs~\cite{esquivel1:02,esquivel2:02,esquivel:03} and for finite slabs using, for example, the transfer matrix formalism~\cite{esquivel:01,esquivel3:02}. In addition, one observes that the Lifshitz formula is symmetric under the change of index $j= 1 \to 2$, it means that the force is indistinguishable under the change of planes which is natural given the symmetry of the system. 

The force between a sphere and a plane made of arbitrary dielectric materials is commonly calculated using the Proximity Theorem approximation, where the Casimir energy $\mathcal{V}(z)$ is found using the Lifshitz formula. As a consequence, the force between a sphere and a plane is indistinguishable if the sphere is made of a material A and the plane is made of B, or if the sphere is made of B and the plane is made of A. Furthermore, the Proximity Theorem assumes that the proper electromagnetic modes of the sphere and a plane are the same as those for two parallel planes. Such that, the Proximity Theorem approach neglects the pressure radiation coming from the quantum fluctuations on the opposite side of the sphere. In conclusion, in applying the Proximity Theorem one is assuming that the geometry of the system is quite irrelevant to find the Casimir force.

In this paper, we show that the geometry of the system is important. By calculating the zero-point energy of the interacting surface plasmons of a sphere-plane configuration, we obtain that the force between dissimilar materials differ if the sphere is made of a material A and the plane is made of B, and vice versa. We find a general formula to calculate the energy and Casimir force differences between arbitrary materials. We study the specific case of aluminum and gold. We restrict ourself to the non-retarded limit, this means, to the case of small particles at small distances. 

\section{zero-point energy of the interacting surface plasmons}

\subsection{Parallel planes}
In 1968, van Kampen and collaborators~\cite{kampen} showed that the Lifshitz formula in the nonretarded limit is obtained from the zero-point energy of the interacting surface plasmons of the planes. Later, Gerlach~\cite{gerlach} did an extension showing that also in the retarded limit the Lifshitz formula is obtained from the zero-point energy of the interacting surface plasmons. A comprehensive derivation of the Lifshitz formula using the surface modes is also found in Ref.~\onlinecite{milonni}. 

In the case of two parallel planes, in the absence of retardation, the proper electromagnetic modes satisfy the following expression~\cite{kampen} 
\begin{equation}
\left[ \frac{\epsilon (\omega )+1}{\epsilon (\omega )-1}\right] ^{2}\exp
[2kz]-1=0  \label{3}
\end{equation}
for any value of $0\leqslant k\leqslant \infty $. The roots $\omega_{i}(k) $ are identified as the oscillation frequencies of the surface plasmons. To find the proper modes, it is necessary to choose a model for the dielectric function of the planes. To illustrate the procedure for metallic planes, let us employ the Drude model 
\begin{equation}
\epsilon (\omega )=1- \frac{\omega_p^2}{\omega(\omega + i/\tau)}, \label{drude}
\end{equation}
with $\omega_p$ the plasma frequency and $\tau$ the relaxation time. Substituting Eq.~(\ref{drude}) in Eq.~(\ref{3}), and taking into account that for most materials $ 1 \gg (\tau\omega_p)^{-1}$, we find two proper modes that, at large distances ($kd\gg 1$), are given
\begin{equation}
\omega_{\pm }\approx \frac{\omega_p}{\sqrt{2}}\left( 1\pm \frac{1}{2}\exp
[-kd]-\frac{1}{8}\exp [-2kd] \pm \cdots \right), 
\end{equation}
and in the limit $kd\rightarrow \infty $, we recover $\omega_\pm = \omega_p/\sqrt{2}$ which is the frequency of the surface plasmon of each half space. The zero-point energy per unit area will be given by
\begin{equation}
{\mathcal E}_{0}(z) =\frac{1}{2} \frac{\hbar \omega _{p}}{\sqrt{2}}2\pi \int_{0}^{\infty } \sum_i \omega_{i}(k) kdk , 
\end{equation}
Now, taking the difference of the zero-point energy when the planes are at a distance $z$, and when they are at infinite, we obtain the interaction energy of the system per unit area, like 
\begin{equation}
{\mathcal V}(z) = -\frac{\hbar \omega _{p}}{\sqrt{2}} \frac{\pi}{16z^2}.
\end{equation}
which depends on the energy of the surface plasmon of a half space and the distance between planes.
 
\subsection{Sphere and plane: dipolar approximation}
Now, let us consider the case of a sphere of radius $a$ and dielectric function $\epsilon_s(\omega)$, such that, it is at a minimum distance $z$ from a plane of dielectric function $\epsilon_p(\omega)$, as shown in Fig.~1. The quantum vacuum fluctuations will induce a charge distribution on the sphere which also induces a charge distribution in the plane. Then, the induced $lm$-th multipolar moment on the sphere is given by~\cite{claro}  
\begin{equation}
Q_{lm}(\omega)  =  \alpha_{lm} (\omega)\left[ V_{lm}^{\rm vac}(\omega) + V_{lm}^{\rm sub}(\omega) \right], \label{qlm}
\end{equation}
where $V_{lm}^{\rm vac}(\omega)$ is the field associated to the quantum vacuum fluctuations at the zero-point energy, $V_{lm}^{\rm sub}(\omega)$ is the induced field due to the presence of the plane, and $\alpha_{lm}(\omega)$ is the $lm$-th polarizability of the sphere. To show the procedure in detail, first we work in the dipolar approximation, i.e., when $l=1$. In section III, we will calculate the proper modes exactly, taking into account all the high-multipolar charge distributions.

\begin{figure}[tbh]
\centerline {
}
\caption{Schematic model of the sphere-plane system.}\label{f1}
\end{figure}

Using the method of images, the a relation between the dipole moment on the the sphere ${\vec p}_{\rm s}(\omega)$, and the induced dipole moment on the plane ${\vec p}_{\rm p}(\omega)$, is
\begin{equation}
{\vec p}_{\rm p}(\omega) = \frac{1 - \epsilon_p(\omega)} { 1 + \epsilon_p (\omega)} {\mathbb M} \cdot {\vec p}_{\rm s}(\omega). \label{psub}
\end{equation}
Here, ${\mathbb M}= (-1, -1, 1)$ is a diagonal matrix whose elements depend on the choice of the coordinate system. From Eq.~(\ref{qlm}), one finds that the total induced dipole moment on the sphere is
\begin{equation}
{\vec p}_{\rm s}(\omega) =  \alpha(\omega) \left[ {\vec E}^{\rm vac} (\omega) + {\mathbb T} \cdot {\vec p}_{\rm p}(\omega) \right], \label{psph}
\end{equation}
where the interaction between dipoles is coupled by the dipole-dipole interaction tensor 
\[\mathbb T= (3 {\vec r} {\vec r} - r^2{\mathbb I})/r^5.\] 
Here, ${\mathbb I}$ is the identity matrix, and ${\vec r}$ is the vector between the centers of the sphere and the dipole-charge distribution on the plane. From  Fig.~1, we find that ${\vec r}=(0,0,2(z+a))$, such that, ${\mathbb M}\cdot {\mathbb T}= (-1/r^3, -1/r^3, -2/r^3)$ is also a diagonal matrix. Substituting Eq.(\ref{psub}) in Eq.(\ref{psph}), one finds 
\begin{equation} 
\left[\frac{1}{\alpha(\omega)}\mathbb I + f_c (\omega) {\mathbb M}\cdot {\mathbb T} \right] \cdot {\vec p}_{\rm s} (\omega) = \mathbb G(\omega) \cdot {\vec p}_{\rm s} (\omega)= {\vec E}^{\rm vac} (\omega), \label{q2}
\end{equation}
where $\mathbb G (\omega)$ is a diagonal matrix with 
\[f_c (\omega)= [ 1 - \epsilon_{\rm p}(\omega)]/[  1 + \epsilon_{\rm p}(\omega)].\] 
Multiplying Eq.~(\ref{q2}) by $a^3$, one finds that each term of $\mathbb G(\omega)$ is dimensionless and has two parts: the left-hand side which is only associated to the material properties of the sphere through its polarizability $1/\tilde{\alpha}(\omega) = a^3/\alpha(\omega)$, and the right-hand side which is related to the geometrical properties of the system trough $a$, and $z$. The right-hand side also depends on the material properties of the plane trough the function $f_c(\omega)$.

The eigenfrequencies of the sphere-substrate system must satisfy Eq.~(\ref{q2}), and are independent of the exciting field, in this case ${\vec E}^{\rm vac}(\omega)$. Then, these frequencies can be obtained when the determinant of the matrix in the left-side of Eq.~(\ref{q2}) is equal to zero, $\det \mathbb G (\omega) =0 $. Thus, the eigenfrequencies are given by 
\begin{equation}
\left[\frac{1}{\tilde{\alpha}(\omega)} +\frac{f_c(\omega)a^3 }{[2(z+a)]^{3}}\right]^2 
\left[ \frac{1}{\tilde{\alpha}(\omega)} +\frac{2f_c(\omega) a^3 }{[2(z+a)]^{3}}\right] = 0. \label{ceros}
\end{equation}
Until this point, it is necessary to consider a model for the dielectric function of the sphere to find the proper modes. Again, we illustrate the procedure for  metallic sphere and  plane, using the Drude model of Eq.~(\ref{drude}). Moreover, we use the fact that the polarizability of the sphere, within the dipolar approximation and in the quasi-static limit, is given by 
\[ \alpha(\omega) = a^{3} \frac{\epsilon_{\rm s}(\omega) -1} {\epsilon_{\rm s}(\omega) + 2}. 
\]

\begin{figure}[tbh]
\centerline {
}
\caption{Plot of $1/\tilde{\alpha}(\omega)$ in solid line and $f_c(\omega) [2(1+z/a)]^{-3}$ as a function of $\omega/\omega_p$ for different values of $z/a$, for a gold sphere over a gold plane.} \label{f2}
\end{figure}

In Fig.~\ref{f2}a, we plot, $1/\tilde{\alpha}(\omega)$ and $f_c(\omega) [2(1+z/a)]^{-3}$ as a function of $\omega/\omega_p$ for different values of $z/a$, for a sphere and a plane both made of gold. The proper frequencies of the system are given when the black-solid line and the color-dotted lines intersect each other. We observe that there are two different proper modes for each term on Eq.~(\ref{ceros}), giving a total of six modes for each $z/a$. The proper modes at the left-hand side in Fig.~\ref{f2}a correspond to the surface plasmon of the sphere that we denote with $\omega_+$. These modes are red-shift as the sphere approaches the plane because of the interaction between them. When the separation $z/a$ increases, the modes $\omega_+$ go to $ \omega_p/\sqrt{3}$, which is the surface plasmon of the isolated sphere. The proper modes at the right-hand side correspond to the surface plasmon of the plane and we denote them with $\omega_-$. These modes are blue-shift as the sphere approaches the plane, also because of the interaction between the sphere and the plane. In this case, when the separation $z/a$ increases, the modes $\omega_-$ go to $\omega_p/\sqrt{2}$ which is the surface plasmon of the isolated plane. This behavior of the modes is clearly observed in Fig.~\ref{f2}b, where $\omega_{+}$ and $\omega_{-}$ are plotted as a function of $z/a$. This shows that the the zero-point energy of the sphere-plane system is directly associated to the interacting surface plasmons of the sphere and the plane. 

\section{Exact calculation of the zero-point energy of a sphere above a plane}

In this section we calculate exactly the proper electromagnetic modes of the sphere-plane system, within the non-retarded limit, by including all the high-multipolar interactions. These proper modes are calculated using a Spectral Representation formalism~\cite{ceci,ceci2}, then, we calculate the zero-point energy using Eq.~(\ref{uint}). The Spectral Representation (SR) formalism has the advantage that the contributions of the dielectric properties of the sphere can be separated from the contributions of its geometrical properties. The latter allows to perform a systematic study of the sphere-plane system that may be very helpful for experimentalists. The details of the SR formalism can be found in Refs.~\onlinecite{ceci}~and~\onlinecite{ceci2}, here we only explain it briefly. 

We now calculate the proper modes of the system including all the high-multipolar charge distributions due to quantum vacuum fluctuations. Using the method of images, we find for the $lm$-th multipolar moment on the sphere, that
\begin{equation}
-\sum_{l'm'} \left[\frac{4 \pi \delta_{ll'} \delta_{mm'} }{(2l+1)\alpha_{l'm'}(\omega)} + f_c(\omega) A_{lm}^{l'm'} \right] {{Q}}_{l'm'} (\omega)= V_{lm}^{\rm vac}(\omega),  \label{q3}
\end{equation}
where $A_{lm}^{l'm'}$ is the matrix that couples the interaction between sphere and substrate~\cite{claro}, and $\alpha_{lm}(\omega)$ is the $lm$-th polarizability of the sphere. The expression for the interaction matrix $A_{lm}^{l'm'}$ is given in the Appendix section, and it shows that $A_{lm}^{l'm'}$ depends only on the geometrical properties of the system. If we have a homogeneous sphere, its polarizabilities are independent of the index $m$, and are given by~\cite{claro} 
\begin{equation}
\alpha_l (\omega) = \frac{l[\epsilon_{\rm s}(\omega) - 1]} {l [ \epsilon_{\rm s}(\omega) + 1] + 1} a^{2l+1} = \frac{n_{l0}}{n_{l0} - u(\omega)} a^{2l+1}, \label{alfa}
\end{equation}
where $n_{l0} = l/(2l+1)$, and $u(\omega) = [1 - \epsilon_s(\omega)]^{-1}$. Notice that also in the latter equation the dielectric properties of the sphere are separated from its geometrical properties. By substituting the right-hand side of the above equation, we rewrite Eq.~(\ref{q3}) as:
\begin{eqnarray}
\sum_{l'm'} \Big\{ - u(\omega) \delta_{ll'}\delta_{mm'} + {H}^{l'm'}_{lm} (\omega) \Big\}  \frac{Q_{l'm'}(\omega)} { (l' a^{2l'+1})^{1/2}}  =&& \nonumber \label{multi} \\
 - \frac{(l a^{2l+1})^{1/2}} {4 \pi} V_{lm}^{\rm vac}(\omega), &&
\end{eqnarray} 
where the $lm,l'm'$-th element of the matrix $\mathbb H(\omega)$ is
\begin{equation}
H_{lm}^{l'm'} (\omega)= n_{l'0}  \delta_{ll'}\delta_{mm'} + f_c(\omega) \frac{(a^{l+l'+1})} {4 \pi} (ll')^{1/2} A_{lm}^{l'm'}. \label{h}
\end{equation}
The proper modes are given when the determinant of the quantity between parenthesis in Eq.~(\ref{multi}) is equal to zero, $\det{[- u(\omega) {\mathbb I} + \mathbb H (\omega)]}=0$ or when the product
\begin{equation}
\prod_s [-u(\omega) + n_s(\omega)] =0,
\end{equation}
where $n_s(\omega)$ are the eigenvalues of $\mathbb H(\omega)$ for each $z$. 

The Spectral Representation formalism can be apply in a very simple way. First, we have to choose a substrate to calculate the contrast factor $f_c(\omega)$, with this, we construct the matrix $\mathbb H(\omega)$ for a given $z$, and we diagonalize it to find its eigenvalues $n_s(\omega)$. We can repeat these steps for a set of different values of $z$. Notice that the eigenvalues of $\mathbb H(\omega)$ can be found  without to do any assumption on the dielectric function of the sphere. Once we have the eigenvalues as a function of $z$, finally, we have to consider an explicit form of the dielectric function of the sphere, such that, we calculate the proper electromagnetic modes $\omega_s(\omega)$ trough the relation $u(\omega_s) = n_s(\omega)$. 

To illustrate the procedure, we use again the Drude model for the dielectric function of the sphere, and a plane with a constant dielectric function. In this case, the proper modes are given by
\begin{equation}
\omega_s(z) = -i/\tau + \sqrt{-(1/\tau)^2 + \omega_p^2 n_s(z)}, \label{modos2}
\end{equation}
with $s = (l,m)$, and where the eigenvalues $n_s(z)$, are independent of the frequency. If we consider that $ 1 \gg (\tau\omega_p)^{-1}$, then $\omega_s(z) \approx \omega_p \sqrt{n_s(z)}$. In this case, the  zero-point energy, according with Eq.~(\ref{uint}), is given by
\begin{equation}
{\cal E}(z) = \frac{\hbar \omega_p}{2} \sum_{l,m}\left[ \sqrt{n_{lm}} - \sqrt{n_{l0}} \right],
\end{equation}
where $\hbar \omega_p\sqrt{n_{l0}} = \hbar \omega_p\sqrt{l/(2l+1)}$, are the energies associated  to the multipolar surface plasmons resonances of the isolated sphere. Here, , $l = 1, 2, \dots, L$, where $L$ is the largest order in the multipolar expansion. For $l=1$ we have that $\hbar \omega_p\sqrt{n_{10}}= \hbar \omega_p/\sqrt3$, that is the surface plasmon of the sphere in the dipolar approximation. When $l \gg 1$, we have that the surface plasmons of the sphere due to high-multipolar moments are very similar to the surface plasmons of the plane, because $\hbar \omega_p\sqrt{n_{l0}} \to  \hbar \omega_p/\sqrt2$ when $l \to \infty$. In the presence of the plane, the proper modes of the isolated sphere are modified according with Eq.~(\ref{h}). These proper modes are red-shift always as the sphere approaches the plane, and this shift depends on $z/a$ in a non-monotonic way.  As $z/a \to 0$, more and more multipolar interactions must be taken into account. Since the proper modes are red-shift as $z/a \to 0$, it might be possible that $n_s(z) \sim (\tau\omega_p)^{-2}$. In such case, we will have to consider also temperature effects, because it is possible to excite modes of very low frequencies~\cite{ceci:00}. In this work, we restrict our selves to distances such that $n_s(z) \gg (\tau\omega_p)^{-2}$, temperature effects will be studied elsewhere.

It is possible to find the behavior of the energy as a function of $z/a$ for any Drude's sphere independently of its plasma frequency, like $\tilde{{\cal E}} \equiv {\cal E}/\hbar \omega_p $. Furthermore, the force can be also studied independently of the radius of the sphere, since ${\cal E}(z/a)$, one finds that 
\begin{equation}
{ F}=-\frac{\partial{\cal E}(z/a)}{\partial z} = -\frac{\partial{\cal E}(z/a)}{\partial(z/a)}  \frac{\partial(z/a)}{\partial z}, \label{force}
\end{equation}
and then, we can define also a dimensionless force like $a \tilde{ F} \equiv a {F} / \hbar \omega_p$. In summary, we have shown, for a Drude sphere with $ 1 \gg (\tau\omega_p)^{-1}$, that given the dielectric properties of the substrate the behavior of the energy and force is quite general, showing the potentiality of the Spectral Representation formalism. 

In Fig.~\ref{f3}, we show the dimensionless energy, $\tilde{{\cal E}}$, and force $a\tilde{F}$, for a sphere and a plane both made of gold. We show $\tilde{{\cal E}}$ and $a\tilde{F}$ as a function of $z/a$ when all multipolar interactions are taken into account, as well as only dipolar, and up to quadrupolar interactions are considered. The largest multipolar moment to achieve convergence of the energy at a minimum separation of $z/a=0.1$ is $L = 80$. We observe that $\tilde{{\cal E}}$ and $a\tilde{F}$ are a non-monotonic functions of $z/a$ when  we consider multipolar interaction greater than dipolar ones. Only within the dipolar approximation, $\tilde{{\cal E}}$ and $\tilde{F}a$ behaves like $(z/a)^{-3}$, and $(z/a)^{-4}$ for any $z$, respectively. Let us first analyze the system when only up to quadrupolar interactions are taken into account. In that case,  the energy as well as the force, show three different regions: (i) at large distances, $z > 5a$, only dipolar interactions are important, (ii) when $ 5a > z > 2a$, we found that dipolar-quadrupolar interactions become important and $\tilde{{\cal E}}$ and $\tilde{F}a$ behaves like $(z/a)^{-4}$, and $(z/a)^{-5}$; while (iii) at small distances $ z < 2a$ the quadrupolar-quadrupolar interaction dominates, and $\tilde{{\cal E}}$ and $\tilde{F}a$ behaves like $(z/a)^{-5}$, and $(z/a)^{-6}$, respectively

\begin{figure}[htp]
\centerline {
}
\caption{(a) Energy $\tilde{{\cal E}}$, and (b) force $a\tilde{F}$ as a function of $z/a$, with $L=80$ (dotted line), $L=2$ (solid line), and $L=1$ (dashed line). (c) Force difference, $\vert (\tilde{F}^{LH} - \tilde{F}^{LW})/ \tilde{F}^{LH} \vert$, using  multipolar moments $LH=80$ and $LW=2$ (solid line), $LH=80$ and $LW=1$ (dashed line), finally $LH=2$ and $LW=1$ (dot-dot-dashed line).}\label{f3}
\end{figure}

However, as the sphere approaches the substrate the interaction between high-multipolar moments becomes more and more important. The attractive force suddenly increases more than four orders of magnitude, as compare with the dipolar and quadrupolar approximations. In Fig.~\ref{f3}c we show the difference of the dimensionless force, $\vert (\tilde{F}^{LH} - \tilde{F}^{LW})/ \tilde{F}^{LH} \vert$, as a function of the separation, with $LH$ and $LW$ the highest and lowest-multipolar moments taken into account. We observe that at distances $ z > 2a$ the force can be obtained exactly if up to quadrupolar interactions are considered. We also found that for $z > 7a$ the interaction between the sphere and the substrate can be modeled using the dipolar approximation only, like in the Casimir and Polder model~\cite{casypol}. However, at separations smaller than $z < 2a$, the quadrupolar approximation also fails, and it is necessary to include high-multipolar contributions. For example, at $z=a$ the quadrupolar approximation gives an error of about $10\%$, while the dipolar approximation gives an error of about $40\%$. At smaller distances like $z=a/2$, the quadrupolar approximation gives an error of about $40\%$, while in the dipolar approximation the error is $70\%$. At $z=0.1a$ both approximations give an error larger than $90\%$. To directly compare the dipolar and quadrupolar approximations, we also plot in Fig.~\ref{f3}c the force difference when $LH=2$ and $LW=1$. In this curve, we clearly observe that even at $z=5a$ there are differences between the dipolar and quadrupolar approximations of about $5\%$. 

The large increment of the force at small separations due to high-multipolar effects could explain the physical origin of, for example, the large deviations observed in the deflection of atomic beams by metallic surfaces~\cite{shih}, as well as some instabilities detected in micro and nano devices. Furthermore, to compare experimental data of some experiments~\cite{mohideen,chan,decca} with the Proximity Theorem approximation, it has been needed to make a significant modification to the Casimir force, since they measured a larger attractive force than the one predicted. They attributed the deviations from the Proximity Theorem approximation to the roughness of the surface which tend to increase the attraction force, essentially as a series of the inverse powers of the separation, $z$. These deviations might be attribute also to the interactions between high-multipolar charge distributions due to quantum vacuum fluctuations. In summary, the measurements indicate that dispersive forces between a polarizable atom or a spherical particle and a planar substrate involves more complicated interactions than the simple dipole model of Casimir and Polder or the Proximity Theorem approximation. 

In this paper, we propose a way to elucidate experimentally two important issues related with the Casimir force: (i) the relevance of the geometry via the surface plasmon interactions between bodies, and (ii) the relevance of the high-multipolar interactions in the sphere-plane model. In the next section, we study the case of the casimir force between a sphere and a plane of dissimilar materials that can help us to clarify these important issues. 

\section{Casimir force between dissimilar materials}

 Using the formalism presented in the previous section, we calculate the energy and casimir force for a sphere and a plane of dissimilar materials. To illustrate our results, we choose gold and aluminum materials described by the Drude model of Eq.~(\ref{drude}), with plasma frequencies $\hbar \omega_p =~8.55$~eV, and $15.8$~eV, and relaxation times $(\tau \omega_p)^{-1} = 0.0126$, and $0.04$, respectively.

In Fig.~\ref{f4}, we show the exact calculated energy and force for an aluminum sphere above a gold plane, $\mathcal E_{\rm Al/Au}$ and $aF_{\rm Al/Au}$, both in solid line, and the same for a gold sphere above an aluminum plane, $\mathcal E_{\rm Au/Al}$ and $aF_{\rm Au/Al}$, both in dashed line. In both figures, we find that the differences are small and cannot be directly observed. In Fig.~\ref{f4}c, we show the differences of energy and force given by
\[
\Delta {\mathcal E} = 2\left \vert \frac{{\mathcal E}_{\rm A/B} - \mathcal E_{\rm B/A}}{\mathcal E_{\rm A/B} + \mathcal E_{\rm B/A}} \right\vert, 
\] and
\begin{equation} 
\Delta \mathcal F = 2\left \vert \frac{\mathcal F_{\rm A/B} - \mathcal F_{\rm B/A}}{\mathcal F_{\rm A/B} + \mathcal F_{\rm B/A}} \right\vert,
\end{equation}
as a function of $z/a$, with A=Al, and B=Au. Here, we observe that for these materials the differences of energy and force vary as a function of $z/a$. At very small distances, ($z < 0.5 a$) the difference of energy and force, under the interchanged of materials between the sphere and the plane, is less than $3\%$. Then, the difference suddenly increases with the minimum separation, and at $z > a$, the difference of energy and force have almost reached their maximum. For these particular materials, the maximum difference in energy and force is about $6\%$, independently of the radius of the sphere. 

\begin{figure}[hbt]
\centerline {
}
\caption{Exact (a) energy and (b) Casimir force as a function of $z/a$, for a sphere made of Al over an Au plane (solid line), and for a sphere made of Au over an Al plane (dashed line). (c) Energy (solid line) and force (dashed line) differences between a sphere made of Al over an Au plane , and for a sphere made of Au over an Al plane.} \label{f4}
\end{figure}

In the previous section, we show that for a minimum separation larger than $2a$, the force can be described with quadrupolar interactions between the sphere and plane. This means that for $z> 2a$ the energy and force are monotonic functions of $z$. This conclusion can be easily deduce from Fig.~\ref{f3}c. We also show in section~II, that  the Casimir energy is given by the interaction of the surface plasmon of the sphere and  plane. In Fig.~\ref{f2}b, we observe that the proper modes of the system for $z > a$ are approximately given by $\hbar \omega_p/ \sqrt{2}$ and $\hbar \omega_p/ \sqrt{3}$, when $L=1$. These yield to conclude that for $z > a$, the energy and force differences are quiet independent of $z$, and are like
\begin{eqnarray} 
\Delta {\mathcal E} \simeq \Delta {\mathcal F} &\simeq & 2 \left \vert \frac{\left(\frac{\hbar \omega_p^{\rm A}}{\sqrt{3}} +  \frac{\hbar \omega_p^{\rm B}}{\sqrt{2}} \right) - \left(\frac{\hbar \omega_p^{\rm B}}{\sqrt{3}} +  \frac{\hbar \omega_p^{\rm A}}{\sqrt{2}} \right)}{\left(\frac{\hbar \omega_p^{\rm A}}{\sqrt{3}} +  \frac{\hbar \omega_p^{\rm B}}{\sqrt{2}} \right) + \left(\frac{\hbar \omega_p^{\rm B}}{\sqrt{3}} +  \frac{\hbar \omega_p^{\rm A}}{\sqrt{2}} \right)} \right \vert, \nonumber \\
& = & 0.202 \left \vert \frac{\hbar \omega_p^{\rm A} -\hbar \omega_p^{\rm B}}{\hbar \omega_p^{\rm A} + \hbar \omega_p^{\rm B}} \right\vert .
\end{eqnarray}
For the case of A=Al, and B=Au, we find that $\Delta {\mathcal F} \simeq  6.0 \% $, which is the value shown in Fig.~\ref{f4}c when $z > 2a$. Then, as the materials become more dissimilar the difference of energy and force become larger, for example, for potassium and aluminum we would find that $\Delta {\mathcal F} \simeq 12.4\%$, while for gold and copper, $\Delta {\mathcal F} \simeq 2.4\%$. When $z <a$ the difference decreases as $z$ also does. As we see previously, as the sphere approaches the plane more and more multipolar interactions become important in the calculation of the Casimir force. The energy associate to the interacting high-multipolar surface plasmons is proportional to $ \hbar \omega_p \sqrt{l/(2l+1)}$, and when $l$ increases, $\hbar \omega_p \sqrt{l/(2l+1)} \to \hbar \omega_p / \sqrt{2}$, the energy associate to the surface plasmon of a semi-infinite plane. Therefore, as the sphere approaches the plane, the interaction between them resembles the interaction between a plane and a set of planes, where each one of they contributes to the energy with a factor of $(z/a)^{-n}$, with $3 < n \leqslant 2L+1$ a positive integer, where $L$ is the largest order of the interacting multipolar surface plasmon. In conclusion, as the sphere approaches the plane, the interaction between them is like the interaction between a plane and a set of planes, and not like the interaction between two planes, as it is found in the Proximity Theorem approximation.

\section{Summary}
In this work, we show that the Casimir force between a sphere and a plane can be calculated through the energy obtained from their interacting surface plasmons. Our result is in agreement with the work of van Kampen and collaborators~\cite{kampen}, and Gerlach~\cite{gerlach} that showed that the Casimir force between parallel planes can be obtained from the zero-point energy of the interacting surface plasmons of the planes. Then, we showed that the Casimir force of a sphere made of a given material A and a plane made of a given material B, is different from the case when the sphere is made of B, and the plane is made of A. We obtained a formula to estimate the force difference for Drude-like materials. We found that the difference depends on (i) the plasma frequency of the materials, and (ii) the distance of separation between sphere and plane. We also found that as the sphere approaches the plane, the force difference becomes smaller and resembles the interaction of a plane and a set of planes. 

\section*{Acknowledgments}
We would like to thank to Prof. Rub\'en G. Barrera for helpful discussions. This work has been partly financed by CONACyT grant No.~36651-E and by DGAPA-UNAM grant No.~IN104201.
 
\begin{widetext}
\appendix*
\section*{Appendix}
The tensor that couples the interaction between the sphere and the plane is~\cite{claro}
\begin{equation}
A_{lm}^{l^{^{\prime }}m^{^{\prime }}} = \frac{ Y^{m-m^{^{\prime }}}_{l+l^{^{\prime }}} (\theta, \varphi)} {r^{l+l^{^{\prime }}+1} } 
\left[ \frac{(4\pi)^3 (l+l^{^{\prime }}+m-m^{^{\prime }})!(l+l^{^{\prime }}-m+m^{^{\prime }})!} {(2l+1) (2l^{^{\prime}}+1)(2l+2l^{^{\prime }}+1)(l+m)!(l-m)!(l^{^{\prime }}+m^{^{\prime}})! (l^{^{\prime }}-m^{^{\prime }})!} \right]^{1/2}
\end{equation} 
where $Y_{lm}(\theta, \varphi)$ are the spherical harmonics. In spherical coordinates ${\vec r}= (2(z+a), 0,0)$, then, the spherical harmonics reduce to~\cite{arfken}
\begin{equation}
Y^{m-m^{^{\prime }}}_{l+l^{^{\prime }}} (0, \varphi) = \left[ \frac{2(l+l^{^{\prime }}) +1}{4 \pi} \right]^{1/2} \delta_{m-m^{^{\prime }}, 0}.
\end{equation}
This means that $m=m^{'}$ and multipolar moments with different azimuthal charge distribution are not able to couple or interact among them. This yields to
\begin{equation}
A_{lm}^{l^{^{\prime }}m^{^{\prime }}} = \frac{4 \pi}{[2(z+a)]^{l+l'+1}} \left[ \frac{ ((l+l')!)^2}{(2l+1) (2l^{^{\prime}}+1) (l+m)! (l-m)! (l^{^{\prime }}+m)! (l^{^{\prime }}-m)!} \right]^{1/2}, 
\end{equation}
which is a symmetric matrix. 
\end{widetext}

\bibliography{disimilar}

%\newpage

%\begin{figure}[h]
%\centerline {
%%\includegraphics[width=3.15in]{fig1}
%}
%\caption{Schematic model of the sphere-plane system.}\label{f1}
%\end{figure}

%\begin{figure}[h]
%\centerline {
%%\includegraphics[width=3.1in]{fig2}
%}
%\caption{Plot of $1/\tilde{\alpha}(\omega)$ in solid line and $f_c(\omega) [2(1+z/a)]^{-3}$ as a function of $\omega/\omega_p$ for different values of $z/a$, for a gold sphere over a gold plane.} \label{f2}
%\end{figure}

%\begin{figure}[h]
%\centerline {
%%\includegraphics[width=3.45in]{fig3}
%}
%\caption{(a) Energy $\tilde{{\cal E}}$, and (b) force $a\tilde{F}$ as a function of $z/a$, with $L=80$ (dotted line), $L=2$ (solid line), and $L=1$ (dashed line). (c) Force difference, $\vert (\tilde{F}^{LH} - \tilde{F}^{LW})/ \tilde{F}^{LH} \vert$, using  multipolar moments $LH=80$ and $LW=2$ (solid line), $LH=80$ and $LW=1$ (dashed line), finally $LH=2$ and $LW=1$ (dot-dot-dashed line).}\label{f3}
%\end{figure}

%\begin{figure}[h]
%\centerline {
%%\includegraphics[width=3.45in]{fig4}
%}
%\caption{Exact (a) energy and (b) Casimir force as a function of $z/a$, for a sphere made of Al over an Au plane (solid line), and for a sphere made of Au over an Al plane (dashed line). (c) Energy (solid line) and force (dashed line) differences between a sphere made of Al over an Au plane , and for a sphere made of Au over an Al plane.} \label{f4}
%\end{figure}
%
\end{document}